\documentclass[twocolumn,aps,noshowpacs,showkeys]{revtex4}

\usepackage{graphicx}
\renewcommand{\epsilon}{\varepsilon}

\newcommand{\figuresize}{0.35}

\begin{document}

\title{Reversible Gel Formation of Triblock Copolymers Studied by Molecular
  Dynamics Simulation}

\author{Lei Guo}
\author{Erik Luijten}
\email[Corresponding author; e-mail: ]{luijten@uiuc.edu}
\affiliation{Department of Materials Science and Engineering,\\
  University of Illinois at Urbana-Champaign, 1304 West Green Street, Urbana,
  Illinois 61801}

\date{September 1, 2004; revised version October 12, 2004}

\begin{abstract}
  Molecular dynamics simulations have been employed to study the formation of a
  physical (thermoreversible) gel by amphiphilic A-B-A triblock copolymers in
  aqueous solution. In order to mimic the structure of hydrogel-forming
  polypeptides employed in experiments [W.A. Petka \emph{et al.}, Science
  \textbf{281}, 389 (1998)], the endblocks of the polymer chains are modeled as
  hydrophobic rods representing the alpha-helical part of the polypeptides
  whereas the central B-block is hydrophilic and semi-flexible. We have
  determined structural properties, such as the hydrophobic cluster-size
  distribution function, the geometric percolation point and pair correlation
  functions, and related these to the dynamical properties of the system.  Upon
  decrease of the temperature, a network structure is formed in which bundles
  of endblocks act as network junctions. Both at short and medium distances an
  increased ordering is observed, as characterized by the pair correlation
  function.  Micelle formation and the corresponding onset of geometric
  percolation induce a strong change in dynamical quantities, e.g., in the
  diffusion constant and the viscosity, and causes the system to deviate from
  the Stokes--Einstein relation. The dynamical properties show a temperature
  dependence that is strongly reminiscent of the behavior of glass-forming
  liquids.  The appearance of a plateau in the stress autocorrelation function
  suggests that the system starts to exhibit a solid-like response to applied
  stress once the network structure has been formed, although the actual
  sol--gel transition occurs only at a considerably lower temperature.
\end{abstract}


\maketitle

\section{Introduction}

Solutions of polymers with attractive groups (associating polymers) exhibit a
wide range of rheological properties that can be controlled through variation
of temperature and concentration. These materials enjoy applications ranging
from viscosity modifiers in food or oil recovery to adhesives and coatings (see
Ref.~\cite{rubinstein99} and references therein).  Under certain conditions,
the attractive groups of the polymers associate to form a network and the
system undergoes a sol--gel transition.  The physical bonds between the
attractive groups are reversible and, depending on their strength, can break
and reform frequently on experimental time scales.  The properties of these
so-called \emph{weak} or physical gels differ markedly from chemical gels in
which the polymers are interconnected through covalent bonds.  Compared to
chemical gels, the current understanding of physical gelation is still limited
and even controversial~\cite{rubinstein99}. Scenarios for thermoreversible
gelation include the possibility of \emph{discontinuous} gelation, in which the
gelation is accompanied by sol--gel phase separation, and \emph{continuous}
gelation. The theoretical treatments of Tanaka and
Stockmayer~\cite{tanaka89a,tanaka89b,stockmayer91,tanaka94} predict that
continuous gelation is a thermodynamic phase transition, whereas Semenov and
Rubinstein arrive at the opposite conclusion~\cite{semenov98}.

Simulations can provide specific information that is not easily obtained
otherwise. On the one hand, the microscopic structure of physical gels is
difficult to determine experimentally, whereas it can be directly accessed
(within the limitations of the model used) by means of simulations. On the
other hand, the numerical calculations permit testing of theoretical hypotheses
and approximations. Two models are widely used to study associating polymers.
In the first category of models, used in the above-mentioned theories, there
are many association sites (``stickers'') distributed along the polymer chain.
This model was studied extensively in early (off-lattice) Monte Carlo
simulations by Groot and Agterof~\cite{groot94a,groot94b,groot95}. Kumar and
Panagiotopoulos~\cite{kumar99} have investigated the thermodynamic properties
of a lattice-based version of this model by Monte Carlo simulations and did not
find any indication that gelation is a thermodynamic phase transition. More
recently, it has been observed that the dynamical properties of this model are
similar to those of weak glass formers, in which the diffusion coefficient is
described by an expression with a Vogel--Fulcher form~\cite{kumar01}.

The second category of models consists of \emph{telechelic} chains, in which
the associative sites are located at both chain ends and typically represented
by a single monomer (see Ref.~\cite{larson98} for a concise overview). At low
concentrations, these chains have been predicted to form flower-like
micelles~\cite{semenov95}. Simulations have indeed confirmed
this~\cite{khalatur96,bedrov02} and found that the associative groups are
located in the core of the micelles and the non-associative groups in the
corona.  At higher concentrations, micelles can be connected by ``bridging''
polymers for which both associative endgroups belong to different micelles,
leading to the formation of a micellar gel.  The dynamics in such a system are
governed by the hopping rate of the associative groups between different
micelles~\cite{semenov95}. Simulations have indeed found that the diffusion
properties of such solutions can be described by an Arrhenius
law~\cite{bedrov02}, as predicted by Tanaka and Edwards~\cite{tanaka92b}.  If
the polymer chains are less flexible, qualitative structural changes occur, as
intra-chain pairing is suppressed and the formation of flower-like micelles
becomes energetically unfavorable. This promotes the formation of a network
structure at low polymer concentrations~\cite{khalatur99}. However, the
dynamical properties of such solutions of semiflexible telechelic chains seem
not to have been investigated.

Recently, Petka \emph{et al.}~\cite{petka98} have used genetic engineering
techniques to create artificial proteins consisting of a hydrophilic group
flanked by two stiff hydrophobic blocks.  This triblock copolymer was found to
exhibit gelation in response to variation of pH or temperature. Its
significance lies in the possibility to independently tune the strength of the
endgroup attractions that are responsible for gelation and the solvent
retention capability of the chains, which is essential for the formation of a
swollen gel.  However, the actual structure of the hydrogel, which is formed at
low polymer concentrations, could only be conjectured. Motivated by the
experimental findings, we have employed molecular dynamics simulations to
investigate the dynamic and structural properties of a solution of triblock
copolymers that can be viewed as a greatly simplified, coarse-grained model of
the artificial proteins. This model evidently does not capture all relevant
properties of the experimental systems, but rather should be viewed as a first
attempt to determine the generic properties of a solution of triblock
copolymers with two stiff endgroups.

\section{Model and simulational details}

In order to study the gelation of triblock copolymers we employ molecular
dynamics simulations, using the DL\_POLY\_2 code~\cite{smith96}.  The polymers
have an A-B-A structure, where the A-blocks are rigid hydrophobic rods and the
B-block is hydrophilic and semiflexible. In our coarse-grained model, the
solvent is modeled implicitly and each copolymer block is composed of spherical
units (``monomers'') that represent an effective segment. The total length of
each chain is set to 15 units, consisting of three A-monomers per hydrophobic
block and nine B-monomers in the hydrophilic block.  This choice is mostly
based upon practical considerations.  A minimum of three units is required to
represent a rod-like endblock, whereas a longer central block would pose
equilibration problems, given the computationally accessible time scales.
Monomers of type~A interact via an attractive Lennard-Jones potential,
\begin{equation}
U_{\rm AA}=4\epsilon_{\rm AA}\left[ \left(\frac{\sigma_{\rm
AA}}{r}\right)^{12}- \left(\frac{\sigma_{\rm AA}}{r} \right)^6\right] \;,
\label{eq:lj}
\end{equation}
whereas the interactions between monomers of type~B and the interactions
between unlike pairs are purely repulsive,
\begin{eqnarray}
U_{\rm BB} &=& 4\epsilon_{\rm BB}\left(\frac{\sigma_{\rm BB}}{r}\right)^{12}
\nonumber\\ 
U_{\rm AB} &=& 4\epsilon_{\rm AB}\left(\frac{\sigma_{\rm AB}}{r}\right)^{12}
\;.
\label{eq:repulsion}
\end{eqnarray}
We set $\epsilon_{\rm AA}=\epsilon_{\rm BB}=\epsilon_{\rm AB}=\epsilon$ and
$\sigma_{\rm AA}=\sigma_{\rm BB}=\sigma_{\rm AB}=\sigma$ and cut off all
interactions at $2.5\sigma$. In order to express our results in reduced units
we use $\epsilon$ and $\sigma$ as units of energy and length, respectively.
The reduced coupling, or inverse reduced temperature, $J \equiv \epsilon/k_{\rm
B} T$ is varied between 1 and 2 in the simulations.  The semiflexible character
of the hydrophilic block is controlled by a harmonic angle-dependent potential,
\begin{equation}
U_{\theta}=\frac{1}{2}k_{\theta}(\theta-\theta_0)^2 \;,
\end{equation}
where $k_{\theta}=10\epsilon/{\rm degree}^2$ and $\theta_0=175^\circ$.  The
value of $\theta_0$ was chosen in accordance with the model proposed in
Ref.~\cite{khalatur99}. In combination with the large value for $k_\theta$ it
causes the chains to adopt an extended structure (but without having the
rod-like structure that would be obtained for $\theta_0=180^\circ$), and
gelation is anticipated to occur at relatively low polymer concentrations. The
$\alpha$-helical structure in the artificial proteins~\cite{petka98} is
mimicked by making the endblocks fully rigid. All monomer units within a chain
are connected via a harmonic bond potential
\begin{equation}
U_{\rm bond} =  \frac{1}{2}k_{\rm bond}(r-r_0)^2 \;,
\label{eq:bond}  
\end{equation}
where $k_{\rm bond} = 170\epsilon/\sigma^2$ and $r_0=1.30\sigma$.

The simulations are performed in the canonical ($NVT$) ensemble, in a cubic box
of linear dimension~$L=39\sigma$ with periodic boundary conditions. The total
number of chains equals $N=216$, corresponding to a monomeric packing fraction
of only~$0.029$, i.e., roughly twice the overlap threshold. The temperature is
controlled by means of the Nos\'{e}-Hoover thermostat~\cite{frenkel-smit2}. The
equations of motion are integrated using a ``leap-frog'' Verlet
scheme~\cite{allentildesley87}, with a time step (in reduced units) $\Delta t =
0.00287$.  In all runs, the system is first equilibrated for four million
steps; for some low temperatures, even longer equilibration periods are used.
Subsequently, 40 million time steps are carried out for high temperatures and
200 million time steps for low temperatures.  After the equilibration period,
the configuration of the system is recorded every 1000 time steps for analysis
of structural (e.g., chain conformations and percolation of the system) and
dynamical (e.g., single-chain diffusion) properties.  In addition, the energies
and the stress tensors are calculated and recorded every 100 time steps for the
calculation of the specific heat and the stress autocorrelation function.

Despite the simplifications made in this coarse-grained model, the required
simulation effort is still appreciable. The total amount of CPU time
corresponds to approximately 2.5 years on a single 2.0GHz Intel Xeon processor.

\section{Simulation results}

\subsection{Structural properties}

\begin{figure}
  \centering
  \scalebox{\figuresize}{\includegraphics[angle=-90]{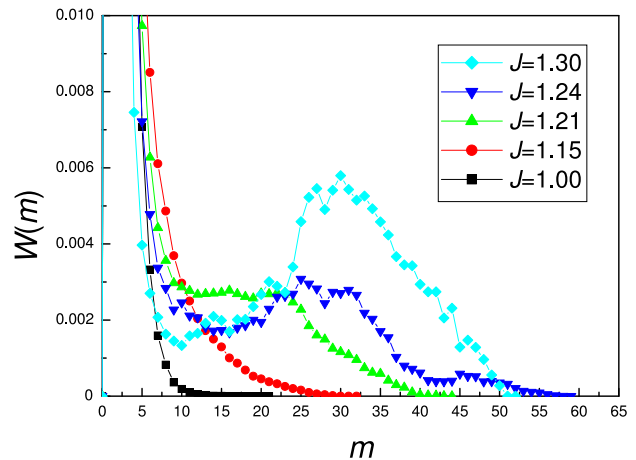}}
  \caption{Bundle-size distribution [probability~$W(m)$ of encountering a
  bundle containing $m$ endblocks] for five representative values of the
  inverse temperature~$J$. The appearance of a secondary peak characterizes the
  formation of bundles in the solution.}
  \label{fig:bundle_dist}
\end{figure}

In order to characterize structural changes that take place in this system upon
variation of the temperature, we employ an approach used in the study of
micelle formation. For each configuration, bundles of endblocks are identified.
An endblock is considered part of a bundle if its center monomer lies within
a distance~$r_{\rm c}$ from the center monomer of an endblock that is already
part of the bundle. Our results turn out to be insensitive to the precise value
of $r_{\rm c} \in [\sigma, 4\sigma]$ and we have chosen $r_{\rm c}=2\sigma$. A
configuration contains $N(m)$ bundles of $m$ endblocks and the bundle-size
distribution is defined as the thermal average~\cite{khalatur99}
\begin{equation}
W(m) = \frac{\langle N(m) \rangle }{\langle \sum_{m}N(m)\rangle} \;.
\end{equation}
Figure~\ref{fig:bundle_dist} shows $W(m)$ for five values of the inverse
temperature~$J$, illustrating the formation of bundles of hydrophobic blocks as
the temperature is lowered.  At high temperature ($J=1.00$), the distribution
function decays monotonically, with a single peak at $m=1$ (isolated
endblocks). This corresponds to a regular solution of chains that are not
associated. For lower temperatures ($J > 1.15$), however, an additional
``shoulder'' appears in the distribution function, which develops into a peak
that increases in height and shifts to larger bundle sizes if the temperature
is further decreased. This signals the formation of bundles in which large
numbers of hydrophobic endblocks participate, resulting in important structural
changes in the solution. The occurrence of an inflection point in $W(m)$ has
been taken as a criterion for the critical micelle point~\cite{khalatur99}.
Here, we associate the appearance of a shoulder in Fig.~\ref{fig:bundle_dist}
with the onset of the bundling process.  The corresponding characteristic
inverse temperature~$J^*$ lies between $1.15$ and~$1.20$.

\begin{figure}
  \centering 
  \scalebox{\figuresize}{\includegraphics[angle=-90]{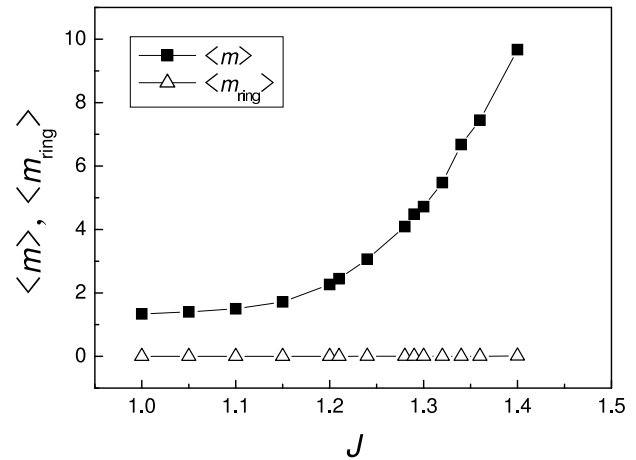}}
  \caption{Comparison between the average number of endblocks per bundle
  (closed squares) and the average number of endblocks per bundle that form a
  loop-like structure (open triangles). Whereas the average bundle size
  increases upon increasing coupling~$J$ (decreasing temperature), the average
  number of loops remains negligibly small. The rigidity of the chains thus
  prevents the formation of ``flower-like'' micelles in the solution. Error
  bars are smaller than the symbol size.}
  \label{fig:ring}
\end{figure}

A similar criterion was employed in Ref.~\cite{bedrov02} [but note that the
quantity $P(m)$ employed in this reference differs from~$W(m)$], and a
comparable temperature dependence was observed for the bundle-size
distribution. There is, however, a marked difference in the morphology of the
bundles. The flexible telechelic polymers studied in Ref.~\cite{bedrov02} form
flower-like micelles in which the chains take the shape of a loop. Both
endgroups of each chain lie in the core of the same micelle and the central
block lies in the corona~\cite{khalatur99}.  For more rigid chains, two effects
occur. The stiff endblocks have a tendency to align inside the bundle, giving
it the appearance of a microcrystalline domain. In addition, the semiflexible
character of the central block prevents the chain from adopting a loop-like
conformation. This second effect is illustrated in Fig.~\ref{fig:ring}, in
which we compare the average number of endblocks per bundle to the average
number of endblocks in a bundle that belong to the same chain.  As shown,
single-chain loops are essentially absent at all temperatures, confirming the
prohibitively large energy penalty incurred by ring formation.

\begin{figure}
  \centering
  \scalebox{\figuresize}{\includegraphics[angle=-90]{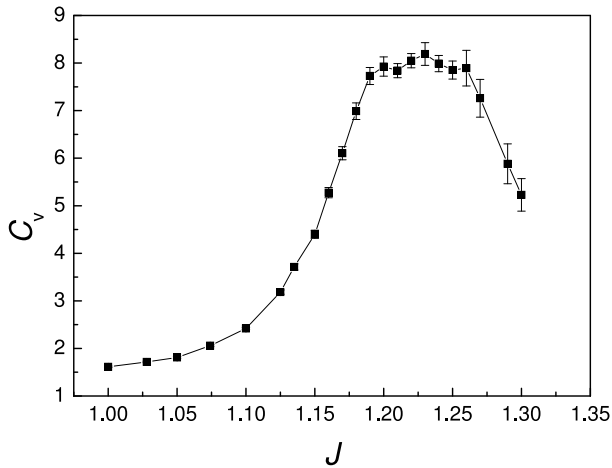}}
  \caption{The specific heat $C_V$ (in reduced units) as a function of inverse
  temperature~$J$. The maximum is indicative of bundle formation in the system.
  The line serves as a guide to the eye.}
  \label{fig:Cv}
\end{figure}

Instead, each chain takes an extended conformation, with both hydrophobic
endgroups participating in different bundles. Thus, even the formation of a
continuous network becomes possible, in which the bundles of endblocks act as
network junction points~\cite{khalatur99}. However, following the experimental
observations in Ref.~\cite{petka98}, we have chosen a markedly lower
concentration than in earlier simulation studies. Before investigating whether
the formation of a network is nevertheless possible, we consider the energetic
aspects of bundling.  As illustrated in Fig.~\ref{fig:Cv}, the specific
heat~$C_V$ exhibits a pronounced but relatively broad maximum around $J=1.23$,
corresponding to the creation of bundles of attractive endblocks. The
specific-heat maximum was found to occur at a temperature \emph{below} the
onset of micelle formation (at $J=J^*$) in Ref.~\cite{bedrov02}. Our data do
not permit us to conclude this unambiguously.  Indeed, for $J \gtrsim 1.20$ the
simulations become almost prohibitively expensive, owing to the slow dynamic
evolution of the system. Thus, the present system does not lend itself well to
the application of finite-size scaling techniques for the determination of the
nature and precise location of the bundling transition. For example, in case of
a continuous phase transition, the height of the specific-heat maximum will
increase (up to corrections to scaling) as $L^{\alpha/\nu} \propto
N^{\alpha/(3\nu)}$. The exponent $\alpha/(3\nu)$ is typically rather small
(e.g., $0.058$ for Ising-type criticality~\cite{ising3d}), so that even
doubling the number of chains would only increase the peak height by an amount
comparable to the statistical accuracy of the data.

\begin{figure}
  \centering
  \scalebox{\figuresize}{\includegraphics[angle=-90]{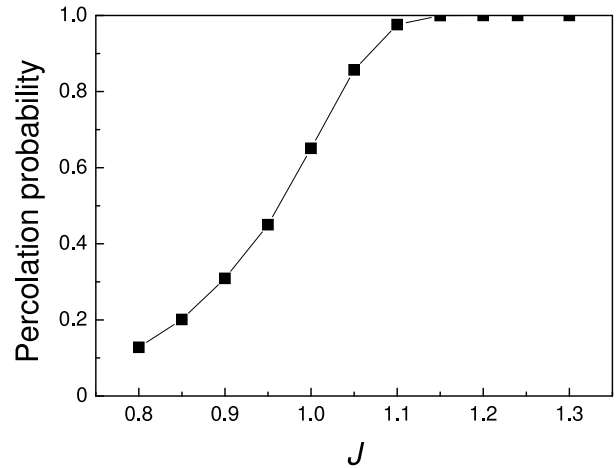}}
  \caption{Percolation probability of a solution of associative telechelic
  polymers at a monomeric packing fraction $\phi = 0.029$, as a function of
  inverse temperature~$J$. Since the percolation probability equals unity for
  $J \gtrsim 1.15$, we consider this to be a measure for the percolation
  threshold.}
  \label{fig:percolation}
\end{figure}

\begin{figure}
  \centering
  \scalebox{0.44}{\includegraphics{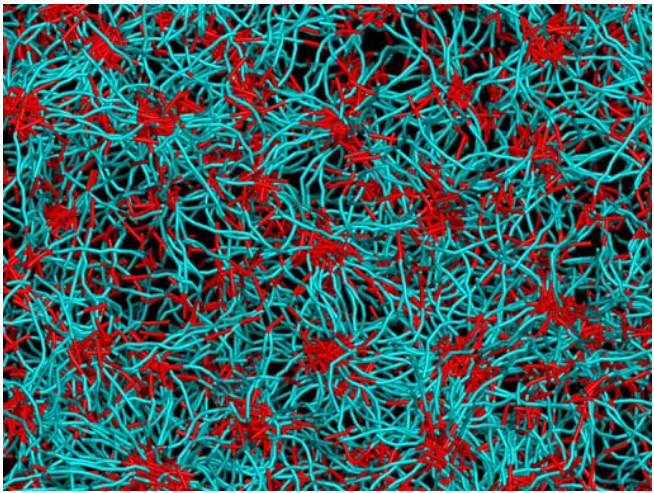}}
  \caption{Snapshot of a simulation at $J=1.30$, for a system of linear size
  $L=78\sigma$ (monomer packing fraction~$0.029$). The image represents
  approximately two-thirds of the simulation box.  The hydrophobic endblocks
  are shown in red and the hydrophilic groups in cyan. The extended structure
  of individual chains as well as the bundling of hydrophobic blocks (cf.\ the
  peak in Fig.~\protect\ref{fig:bundle_dist}) can clearly be seen.}
  \label{fig:snapshot}
\end{figure}

In order to determine whether bundle formation indeed leads to the emergence of
a connected network structure, we consider the percolation probability.
Geometric percolation of the polymer chains in the solution is a necessary
condition for gelation. However, whereas \emph{chemical} gelation coincides
with the occurrence of geometric percolation~\cite{rubinstein03},
\emph{physical} gelation has been suggested to take place only far below the
percolation point~\cite{kumar01}. We consider our polymer solution to be
percolating if a connected path (composed of chains that bridge the bundles of
endblocks) exists between any pair of opposite sides of the simulation cell.
The percolation probability, which is defined as the probability that a
configuration is percolating, is plotted as a function of inverse temperature
in Fig.~\ref{fig:percolation}. The system always percolates for $J \gtrsim
1.15$, i.e., near the characteristic inverse temperature~$J^*$ for bundle
formation. As the percolation probability certainly can exhibit strong
finite-size effects, this determination must only be viewed as an estimate for
the percolation threshold in the thermodynamic limit.  The observation that
$J^*$ and the percolation threshold coincide reaffirms our interpretation that
the telechelic chains become interconnected through bundle formation and form
an spanning network. This behavior appears to differ from what has been
observed for the solution of flexible telechelic chains studied in
Ref.~\cite{bedrov02}, which exhibits a comparable temperature dependence in the
bundle-size distribution but is reported to exhibit geometric percolation at
all temperatures. Figure~\ref{fig:snapshot} shows a typical configuration,
obtained in a simulation performed at $J=1.30$. A network of interconnected
hydrophobic junction points is indeed clearly discernable.

\begin{figure}
  \centering 
  \scalebox{\figuresize}{\includegraphics[angle=-90]{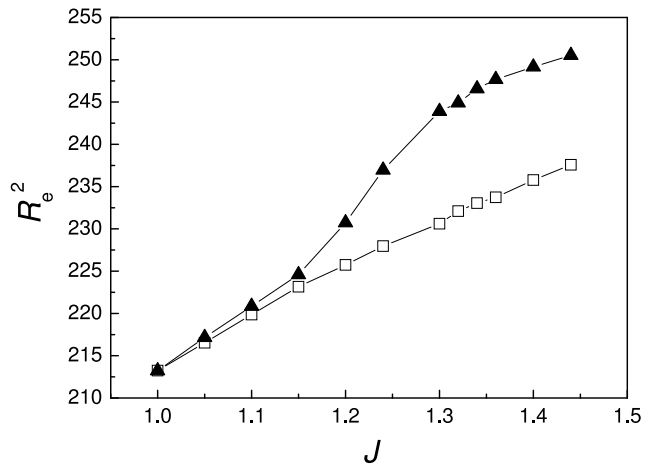}}
  \caption{Square of the end-to-end distance $R_{\rm e}$ of the telechelic
  chains (closed triangles), as a function of inverse temperature~$J$.  For
  comparison, this graph also shows the end-to-end distance for identical
  chains in which the endblocks do \emph{not} possess an attractive interaction
  (open squares). For both chain types, $R_{\rm e}$ increases upon
  increasing~$J$ (decreasing temperature), reflecting the decreasing
  flexibility of the center blocks. However, for $J > J^* \approx 1.15$ the
  telechelic chains clearly exhibit a stronger tendency to adopt an extended
  structure, which is attributed to the formation of an interconnected
  network. Error bars are smaller than the symbol size.}
  \label{fig:Re2}
\end{figure}

The onset of percolation affects the single-chain conformations as well. This
is illustrated by means of the temperature dependence of the end-to-end
distance~$R_{\rm e}$, see Fig.~\ref{fig:Re2}. For comparison, the figure also
includes the end-to-end distance for an identical system in which the monomers
in the endblocks experience a purely repulsive interaction [see
Eq.~(\ref{eq:repulsion})]. Whereas $R_{\rm e}$ increases for both systems as
the temperature is lowered, the end-to-end distance increases more rapidly for
the chains with attractive endblocks than for the purely repulsive chains.
Because of the semiflexible character of the chains, the \emph{relative} change
in $R_{\rm e}$ is only several percent, but nevertheless the effect is clearly
most pronounced for $J \gtrsim 1.15$, i.e., near the percolation
threshold~$J^*$. We ascribe it to the conformational changes induced by the
network formation. Owing to the low polymer concentration, the bundles are
relatively widely separated, forcing the connecting chains to adopt an extended
conformation. This observation is reinforced by considering the spatial
correlations between endblocks.

\begin{figure}
  \centering \scalebox{\figuresize}{\includegraphics[angle=-90]{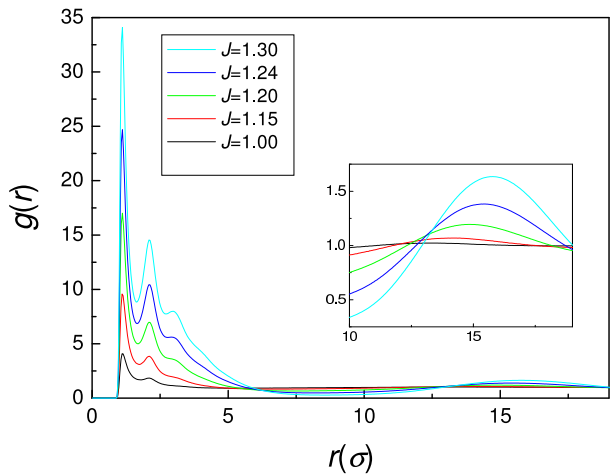}}
  \caption{Hydrophobe--hydrophobe radial distribution function at different
  values of the inverse temperature~$J$. At short distances, the existing peaks
  increase and new peaks emerge as $J$ increases, an indication of the
  alignment of the rigid hydrophobic rods within a bundle toward a
  microcrystalline-like structure. At long distances, a new peak appears at the
  average bundle-to-bundle distance and grows higher at higher $J$, suggesting
  the structured arrangement of the bundles.}
  \label{fig:rdf}
\end{figure}

Indeed, the rod-like structure of the hydrophobic endblocks leads to an
internal structure in the bundles that is absent in the models studied in
Refs.~\cite{khalatur99,bedrov02}. Figure~\ref{fig:rdf} shows the
hydrophobe--hydrophobe radial distribution function~$g(r)$ (calculated from
their center-of-mass separation) at different values of~$J$. As the temperature
is decreased, two distinct features can be identified in this distribution
function. The increasing maxima at short separations, which all lie at
distances within the bundle size (cf.\ Fig.~\ref{fig:bundle_dist}), correspond
to intra-bundle alignment of endblocks.  The emergence of this microcrystalline
morphology can be understood from the fact that in an aligned bundle each
endblock experiences a large number of monomer--monomer interactions with
surrounding endblocks. The cutoff distance employed in the Lennard-Jones
potential~(\ref{eq:lj}) is larger than the maximum distance between monomers on
fully-aligned (close-packed) endblocks, so that even a single pair of rods can
have $9$ pair interactions.  As shown in Fig.~\ref{fig:ring}, the average
number of rods per bundle increases rapidly from approximately~$2$ at $J=1.15$
to almost~$10$ at $J=1.40$, leading to tightly-bonded bundles. It is this
bonding that makes the resulting network resistant to external stress. A second
feature arises in Fig.~\ref{fig:rdf} at lower temperatures. As shown in the
inset, an additional peak appears at a position that roughly coincides with the
calculated average bundle separation, which varies from~$13$ at $J=1.15$ to
$15$ at $J=1.30$. Thus, this peak characterizes the ordered arrangement of the
bundles at low temperatures, and we conclude that the radial distribution
function reflects the simultaneous emergence of both short-range and
medium-range order upon cooling.

\subsection{Dynamical properties}

\begin{figure}
  \centering
  \scalebox{\figuresize}{\includegraphics[angle=-90]{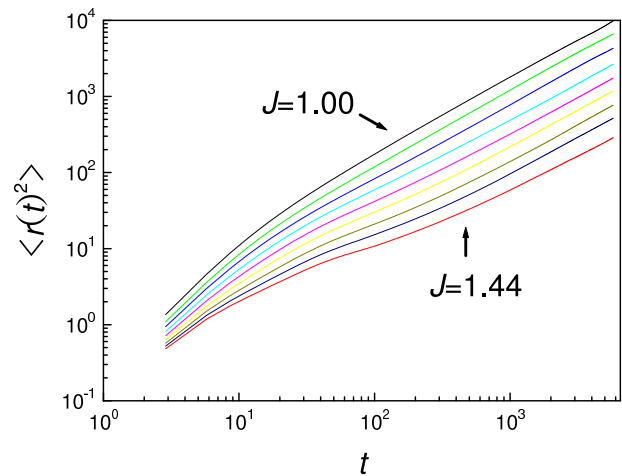}}
  \caption{Mean-square displacement $\langle r(t)\rangle^2$ of the
  single-chain center-of-mass, for inverse temperatures $J=1.00$, $1.15$,
  $1.20$, $1.24$, $1.28$, $1.32$, $1.36$, $1.40$, $1.44$ (only the lowest and
  highest value for $J$ are labeled). In addition to ballistic motion at short
  times and diffusive motion at long times, a slow intermediate regime appears
  at low temperatures.}
  \label{fig:msd}
\end{figure}

\begin{figure}
  \centering
  \scalebox{\figuresize}{\includegraphics[angle=-90]{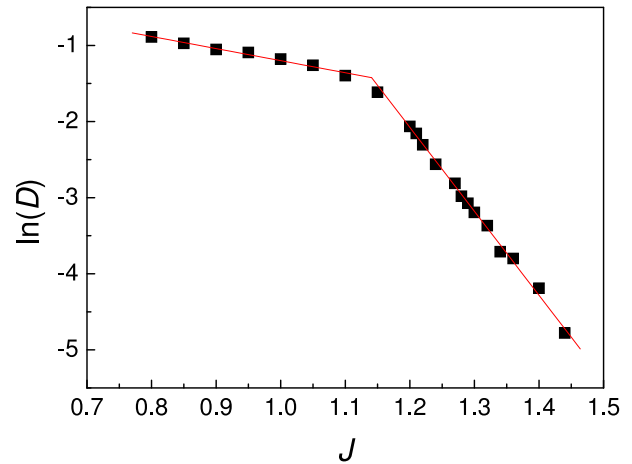}}
  \caption{Diffusion coefficient~$D$ as a function of inverse temperature $J$
  on a log--linear scale. There are two regimes with a different exponential
  dependence on~$J$, which are joined near the percolation point $J^* \approx
  1.15$. While the high-temperature data ($J < J^*$) only permit an approximate
  fit, the diffusion coefficient in the low-temperature regime is clearly well
  described by an Arrhenius law, suggesting that activated processes control
  the relaxation of the system. Error bars are of the order of the symbol size
  or less.}
  \label{fig:diffusion}
\end{figure}

In order to determine whether the structural changes observed in the triblock
copolymer solution indeed correspond to gelation, we consider the dynamical
properties as a function of temperature. Evidently, bundle formation and the
formation of a percolating network are anticipated to have a strong influence
on the diffusion properties of the polymers.  Figure~\ref{fig:msd} shows the
mean-square displacement of the center-of-mass of polymers at different values
for~$J$. At high temperatures, we observe the standard behavior in which the
dynamics cross over from ballistic motion at short times to diffusive motion at
long times. At low temperatures, an intermediate regime appears where the
dynamics are slowed down, indicative of the arrested dynamics resulting from
network formation.  Comparable observations were reported by Kumar and
Douglas~\cite{kumar01} in a Monte Carlo study of a lattice model of an
associating polymer solution and by Bedrov \emph{et al.}~\cite{bedrov02} for
micellar solutions, although it should be noted that in both studies the
polymer concentration was considerably higher than in the current system (which
has $c/c^* \approx 2$) and that in the micellar system the change in dynamic
behavior was not associated with the formation of a network structure. The
dynamic behavior seen in the low-density gel is similar to that found in
glass-forming materials, but the underlying mechanism is different. The
temporary localization of the triblock copolymers is caused by the strong
intra-bundle interactions experienced by the endblocks, rather than by caging
or jamming effects. We also note that the width of the ballistic regime depends
on the simulation model.  If water molecules are included explicitly or
implicitly via a friction coefficient (Brownian dynamics), the ballistic regime
may be rather narrow or even not observable at all. For all investigated
temperatures, the polymers eventually diffuse and we extract the diffusion
coefficient~$D$ from this long-time behavior, see Fig.~\ref{fig:diffusion}. At
low temperatures, a second (narrow) diffusive regime arises for times between
the ballistic and the intermediate regime. This corresponds to intra-domain
motion of the endblocks.

The diffusion coefficient exhibits an exponential dependence on inverse
temperature over the entire temperature range that we investigated, but two
regimes can be discerned, separated near $J^* \approx 1.15$. Since both
percolation and micelle formation occur near this temperature, it is not
possible to uniquely attribute the strong decrease of~$D$ to either of these
two phenomena.  For the ``sticker'' model of Ref.~\cite{kumar01}, in which
association sites are distributed along the polymer chain, the diffusion
coefficient was found to be well described by a Vogel--Fulcher law. In our
system, the diffusive dynamics follow an Arrhenius law, $D\propto {\rm
exp}(-E/k_{\rm B}T)$, where $E$ is the effective activation energy, similar to
strong glass-formers (cf.\ Ref.~\cite{kob99}) and micelle-forming telechelic
polymers~\cite{bedrov02}.  The Arrhenius-type behavior suggests that dynamic
relaxation, which takes place through the exchange of an endblock between two
bundles (network junctions), is controlled by an energetic barrier.  This
barrier has a clear thermodynamic origin, namely the strong attraction between
endblocks. From Fig.~\ref{fig:diffusion}, $E$ is estimated to be approximately
$10\epsilon$ in the low-temperature regime, consistent with our earlier
estimate of the number of interacting monomers in a pair of hydrophobic rods.
Interestingly, the similarity between the dynamic properties observed in this
system and those of the micellar system studied in Ref.~\cite{bedrov02} suggest
that while the semiflexible character of the center blocks and the presence of
rod-like hydrophobic endblocks change the structural properties of the
solution, these differences do not qualitatively affect the dynamic behavior.

Since gelation will be accompanied by a dramatic increase in viscosity~$\eta$,
we compute this quantity by integrating the stress autocorrelation function
$G(t)$~\cite{haile92},
\begin{equation}
\eta = \int_0^{+\infty} G(t) \,dt \;.
\label{eq:visc}
\end{equation}
Here, $G(t)$ is defined as
\begin{equation}
G(t)=\frac{V}{3k_{\rm B}T}\sum_{\rm \alpha \beta}\langle \sigma_{\rm \alpha
  \beta}(t_0)\sigma_{\rm \alpha \beta}(t_0+t)\rangle \;,
\label{eq:stress-corrfn}
\end{equation}
where $V$ is the volume of the system and $\alpha \beta$ assumes the values
$xy$, $yz$, $zx$. The average $\langle \cdots \rangle$ is taken over all time
origins $t_0$. $\sigma_{\rm \alpha \beta}$ is the stress tensor of the
system~\cite{haile92}
\begin{equation}
\sigma_{\rm \alpha \beta} = m\sum_{i=1}^{N}v_{i\alpha}v_{i\beta} +
\frac{1}{2}\sum_{i\ne j}^{N}r_{ij\beta}F_{ij\alpha} \;,
\end{equation}
where $m$ is the monomer mass, $v_{i\alpha}$ is the $\alpha$-component of the
velocity of atom $i$, $r_{ij\beta}$ is the $\beta$-component of the vector
$\mathbf{r}_{ij}$ separating monomers $i$ and~$j$, and $F_{ij\alpha}$ is the
$\alpha$-component of the force exerted on monomer~$i$ by monomer~$j$. The sum
runs over all $N$ monomers. For the calculation of $G(t)$ we employ a fast
Fourier transform~\cite{allentildesley87}, which accelerates the calculation
by several orders of magnitude compared to the direct calculation method.

\begin{figure}
  \centering
  \scalebox{\figuresize}{\includegraphics[angle=-90]{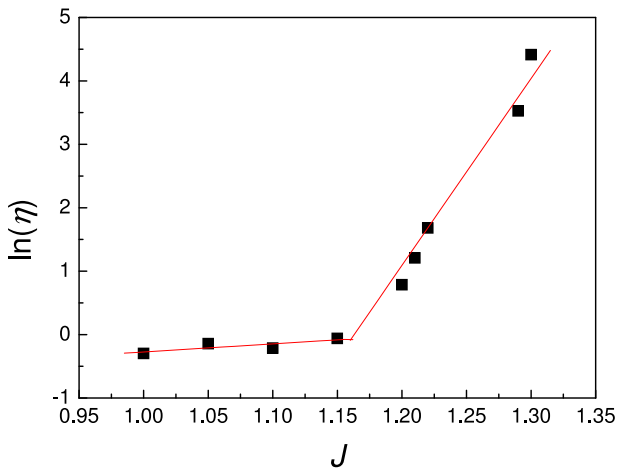}}
  \caption{The viscosity $\eta$ at different couplings on a log--linear
  scale. Similar to $D$, $\eta$ behaves differently in two regions divided at
  the percolation point $J^*=1.15$.  In both regions $\eta$ can be well
  described by an Arrhenius law. Scatter in the data at low temperatures is
  caused by uncertainties in the numerical
  integration~(\protect\ref{eq:visc}).}
  \label{fig:viscosity}
\end{figure}

\begin{figure}
  \centering
  \scalebox{\figuresize}{\includegraphics[angle=-90]{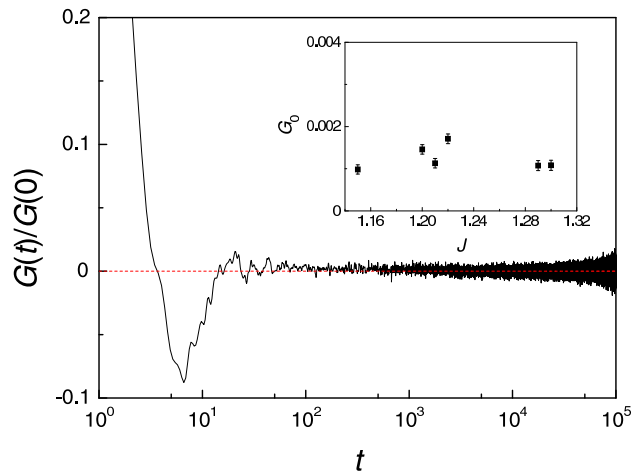}}
  \caption{Normalized stress autocorrelation function $G(t)$ for $J=1.22$. The
  plateau value $G_0$ is small but distinctly nonzero, as shown in the
  inset. In addition, the plateau value does not exhibit a clear temperature
  dependence.}
  \label{fig:autocorr}
\end{figure}

Figure~\ref{fig:viscosity} shows $\eta$ as a function of $J$. Above the
percolation threshold ($J \lesssim 1.15$), the viscosity increases gradually
with decreasing temperature. However, in accordance with the behavior of the
diffusion coefficient, $\eta$ starts to increase rapidly at the onset of
percolation and micelle formation, and is described by an Arrhenius law.
Experimentally, gelation is characterized by the appearance of a plateau in the
stress autocorrelation function $G(t)$.  In our simulations, we observe such a
plateau for all temperatures below the percolation threshold. The plateau
extends to longer times upon decreasing temperature, but eventually $G(t)$
decays to zero.  Since we use the ``atomistic'' (i.e., monomer-based)
representation of the stress tensor~$\sigma_{\alpha \beta}$ in
Eq.~(\ref{eq:stress-corrfn}), the results exhibit relatively large
fluctuations. Figure~\ref{fig:autocorr} shows a representative example. The
plateau value~$G_0$ is small, but clearly nonzero, as shown in the inset. $G_0$
is found to be only weakly dependent on temperature, and no clear trend can be
identified, implying that the rapid increase in~$\eta$
(Fig.~\ref{fig:viscosity}) arises from an increase in relaxation time rather
than from a variation in~$G_0$. The integration~(\ref{eq:visc}) partially
suppresses the statistical fluctuations present in $G(t)$, but the
uncertainties in $\eta$ still reflect the computational challenges, in
particular at low temperatures.  While the strong increase in viscosity follows
unambiguously from Fig.~\ref{fig:viscosity}, we emphasize that the largest
relaxation times (to be discussed below) are still much smaller than the
experimentally observed relaxation times for physical gels, which range from
microseconds to seconds~\cite{rubinstein99}. Thus, only for (computationally
inaccessible) temperatures far below the percolation threshold would the system
investigated here undergo a sol--gel transition.

\begin{figure}
  \centering
  \scalebox{\figuresize}{\includegraphics[angle=-90]{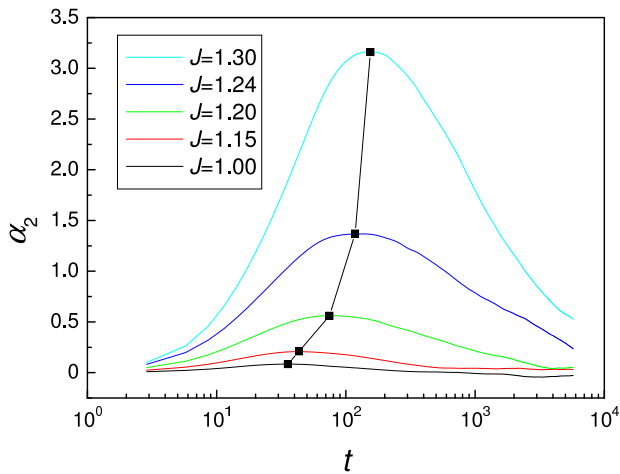}}
  \caption{The non-Gaussian parameter~$\alpha_2$ as a function of time at
  different inverse temperatures~$J$. The deviation of $\alpha_2$ from zero is
  indicative of the heterogeneous dynamics caused by the hopping of endblocks
  between bundles. The peak positions (squares) are used to define a
  characteristic time~$\tau$.}
  \label{fig:nongaussian}
\end{figure}

\begin{figure}
  \centering
  \scalebox{\figuresize}{\includegraphics[angle=-90]{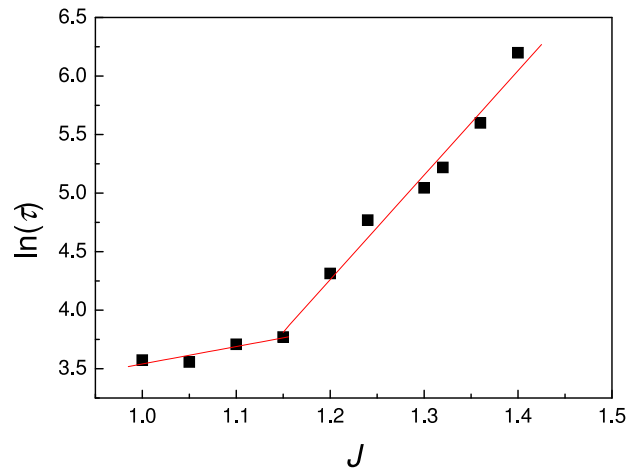}}
  \caption{The characteristic time $\tau$ as extracted from
  Fig.~\protect\ref{fig:nongaussian}, as a function of inverse temperature~$J$
  on a log--linear scale.  Two relaxation regimes can be discerned, separated
  by the percolation threshold $J^* \approx 1.15$.}
  \label{fig:tau}
\end{figure}

Following Ref.~\cite{kumar01}, we employ the non-Gaussian parameter
$\alpha_2$ for the single-chain center-of-mass displacement~\cite{rahman64},
\begin{equation}
\alpha_2 \equiv
  \frac{3}{5} \frac{\langle r(t)^4 \rangle}{\langle r(t)^2 \rangle^2} -1 \;,
\end{equation}
to estimate a characteristic time. This parameter equals zero for both the
ballistic and the diffusive regime. As shown in Fig.~\ref{fig:nongaussian},
$\alpha_2$ increasingly deviates from zero as the temperature is decreased,
reflecting the heterogeneous dynamics resulting from the hopping of endblocks
between bundles. Also the maximum in~$\alpha_2$, from which we extract a
characteristic time~$\tau$, shifts to larger times upon cooling.  At short and
long times, $\alpha_2$ goes to zero, confirming the expected behavior for the
ballistic and diffusive regimes, respectively. At intermediate times, however,
$\alpha_2$ deviates from zero and the deviation increases as the temperature is
lowered. The positive deviation indicates the presence of anomalously fast
chains, very similar to the heterogeneous dynamics observed in glass-forming
liquids~\cite{kob97}.  Figure~\ref{fig:tau} displays $\tau$ as a function of
$J$, on a log--linear scale. We see that the values of $\tau$, which extend up
to approximately $500\tau_0$ for the lowest temperatures, also follow an
Arrhenius law, with a similar change in slope as observed for the diffusion
coefficient and the viscosity. The mean-square displacement corresponding to
$\tau$ (cf.\ Fig.~\ref{fig:msd}) is much smaller than the typical ``hopping
distance'' or bundle separation (cf.\ Fig.~\ref{fig:rdf}), comparable to what
is observed for glassy systems.  Likewise, the observed values of $\tau$ are
much smaller than the relaxation times that follow from the extent of the
plateau in~$G(t)$, which are approximately given by $\eta/G_0(J)$.

\begin{figure}
  \centering
  \scalebox{\figuresize}{\includegraphics[angle=-90]{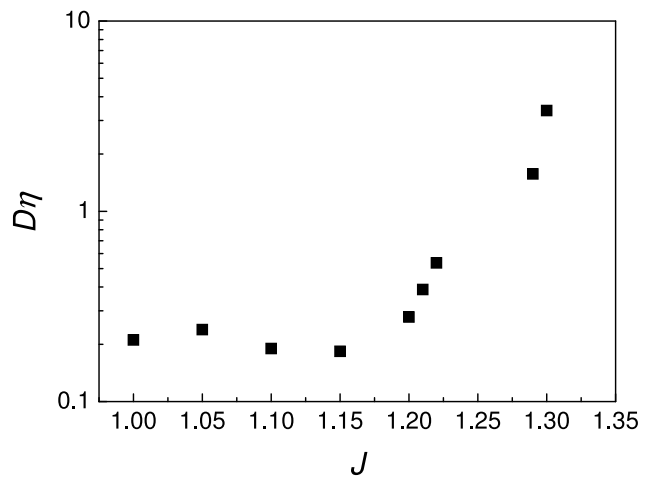}}
  \caption{The product $D\eta$ as a function of $J$. Below $J^*=1.15$, $D\eta$
  varies weakly, consistent with the prediction of the Stokes--Einstein
  equation. Above $J^*$, an order of magnitude increase of $D\eta$ within a
  narrow temperature windows clearly deviates from the Stokes--Einstein
  equation.}
  \label{fig:se}
\end{figure}

Finally, we demonstrate the similarity of our dilute polymer solution to a
glass-forming liquid by considering how well the Stokes--Einstein relation is
obeyed. For a Newtonian fluid, we expect the product of the diffusion
coefficient and the viscosity to obey $D\eta =k_{\rm B}T/4\pi R_{\rm h}$, where
$R_{\rm h}$ is the hydrodynamic radius. The product of $D$ and $\eta$ is
plotted as a function of $J$ in Fig.~\ref{fig:se}. The behavior of the
end-to-end distance (Fig.~\ref{fig:Re2}) suggests that $R_{\rm h}$ will change
only weakly with temperature, and even may increase as the temperature is
lowered. In combination with the linear temperature dependence of the numerator
of the Stokes--Einstein relation (i.e., inverse dependence on~$J$), the steep
\emph{increase} in $D\eta$ for $J \gtrsim J^*$ indicates a clear breakdown of
this relation once micellization and network formation set in.  The diffusion
coefficient is \emph{larger} than would be predicted on the basis of the
viscosity shown in Fig.~\ref{fig:viscosity}. The precise microscopic origin of
this behavior, however, remains to be determined. As shown in the inset of
Fig.~\ref{fig:autocorr}, this breakdown does \emph{not} result from a
temperature dependence in the modulus~$G_0$.

\section{Conclusion}

We have studied the structural and dynamical properties of a solution of
associative A-B-A triblock copolymers. The semiflexible character of the center
(B) blocks, in combination with the attraction between endgroups, allows these
polymers to form a gel at remarkably low concentrations. The molecular dynamics
simulations presented here form a natural extension of earlier work on
semiflexible chains~\cite{khalatur99}, which however only addressed structural
properties. In addition, we observe dynamic effects that bear close resemblance
to those reported in Ref.~\cite{kumar01} for a lattice-based model studied by
Monte Carlo simulations and to those reported in Ref.~\cite{bedrov02} for
micelle formation. However, the solutions in either of these studies had a
significantly higher polymer concentration. Furthermore, the micelle solution
was found to exhibit geometric percolation at all temperatures and the dynamic
changes were linked to the thermodynamic micelle transition.  We observe that
all dynamic changes are correlated with micellization and the simultaneous
emergence of a percolating network of polymer chains, in which bundles of rigid
endblocks act as network junctions. Upon a further decrease in temperature, the
hydrophobic blocks tend to align within a bundle, forming a
microcrystalline-like structure. The resulting strong binding of the chains is
responsible for the mechanical stability of the gel-like network. On a larger
scale, the bundles distribute more regularly at lower temperature, as indicated
by the appearance of peak at the average bundle separation in the
hydrophobe--hydrophobe radial distribution function.

The change in the dynamical behavior of the solution that occurs upon
micellization and network formation is reflected in the diffusion constant, the
viscosity and the maximum in the non-Gaussian parameter. The temperature
dependence of all these properties changes near the percolation threshold and
is well described by an Arrhenius law, similar to what is observed for strong
glass formers. The activation barrier in our system has a clear
\emph{thermodynamic} origin, namely the strong attraction between endblocks
that are part of the same microdomain or ``bundle.'' Thus, there are
similarities with the diffusion of diblock copolymers in a lamellar phase,
which also exhibits an exponential decay with
temperature~\cite{barrat91,lodge95,guenza97b}. Recently, this type of dynamics
has attracted attention in the context of slow dynamics in systems with
frustration-limited domains (see Ref.~\cite{geissler04} and references
therein).

Finally, we note that, while we observe a finite plateau in the stress
autocorrelation function, as would be expected for a gel-forming material, the
dynamics in our systems are still faster than in actual experimental gels.
Thus, the sol--gel transition only occurs at a temperature far below the
percolation threshold.

\begin{acknowledgments}
  Helpful comments by Ken Schweizer are gratefully acknowledged.
  This work is supported by the American Chemical Society Petroleum Research
  Fund under Grant No.\ 38543-G7 and by the National Science Foundation through
  an ITR grant (DMR-03-25939) via the Materials Computation Center at the
  University of Illinois at Urbana-Champaign. Access to computational
  facilities at Oak Ridge National Laboratory (via Grant No.\ CNMS2003-005 at
  the Center for Nanophase Materials Sciences) was supported by the U.S.
  Department of Energy, under contract DE-AC05-00OR22725 with UT-Battelle,
  LLC.
\end{acknowledgments}


\end{document}